\documentclass[a4paper,12pt]{article}
\usepackage[a4paper,margin=2cm]{geometry} 
\usepackage[utf8]{inputenc}
\usepackage[T1]{fontenc}
\usepackage[greek, french]{babel}
\usepackage{graphicx}
\usepackage{caption}
\usepackage{subcaption} % To create subfigures
\usepackage{amsmath}
\usepackage{textcomp}
\usepackage{url}
\usepackage{csquotes}
\usepackage{enumitem}
\usepackage{xcolor}
\usepackage{titlesec}
\usepackage{float} % option [H] des figures
\usepackage{verbatim}
\usepackage{pdfpages}
\usepackage{lmodern}
\usepackage{hyperref}

\title{\fontsize{16pt}{20pt}\selectfont\textbf{Microstates: Do outsiders worth ?}}
\author{Olivier SIRE}
\date{}

\begin{document}

\maketitle

\section{Introduction}

The question of the direction of the  time arrow has sparked extensive debate over the years and will undoubtedly continue to do so. More specifically, its connection to thermodynamic irreversibility has been rigorously scrutinized. While time is not explicitly integrated into the laws of thermodynamics, the second law posits that any process must result in an overall increase in entropy. Jill North \cite{North} argues that within a system-for instance, a gas of randomly distributed particles macroscopically described by Boltzmann’s statistical framework-there may exist microstates exhibiting localized entropy decreases, even as the macro-system’s entropy globally increases. Thus, the temporal evolution toward higher entropy does not preclude the possibility of local microstates with reduced entropy.

In the Big Bang paradigm, as the early universe evolved, minor initial variations in particle densities would have been amplified by gravitational collapse, a self-reinforcing process that recruits additional particles into increasingly dense clumps. Similarly, the Janus model \cite{Janus} proposes that "negative" and "positive" particles attract like charges and repel opposites, triggering the formation of both densely crowded regions and rarefied voids (e.g., the Great Repeller). These mechanisms suggest that systems may host microstates whose properties starkly contrast with the probabilistic macrostate.

It has even been postulated that Earth-or indeed any localized structure like a leaf or a human-could represent a microstate evolving toward lower entropy relative to the universe as a whole. This brief note explores whether cooling systems, such as gases, might generate such microstates and whether these rare, small-scale configurations could give rise to improbably complex structures.

\section{Methods}

To simulate a cooling gas, a simplified computational model was developed. This model features an undersampled universe of constant volume (represented as a 2D surface) where particle density decreases as the universe expands and cools. The simulation is divided into two parts:

    Non-interacting particles: No interactions (e.g., gravitational forces) are modeled.

    Mass-mass interactions: A rudimentary approximation of particle-particle attraction is introduced.

In the first part, the simulation iteratively generates random particle positions using a Mersenne-Twister pseudorandom number generator. The number of particles N ranges logarithmically from 30 to $10^4$, with 100 discrete steps. For each N-particle configuration, 30 random iterations are performed to compute nearest-neighbor distances, the seed being modified at each iteration. The skewness of the resulting spatial distribution is averaged across iterations, and the standard deviation is calculated. The evolution of skewness as a function of particle density is then analyzed. All computations are executed within the R statistical environment.

In the second part, the system is modeled on a 32×32 grid. The temporal evolution of entropy is quantified under a simplified mass-mass attraction scheme (see details below).
\section{Results}
The result of the first simulation is displayed in Fig. 1 which shows the evolution of skewness as the particles density decreases as it happens upon a universe expansion and cooling down. The time arrow is therefore pointing to the right of the graph. Red points are means (n=30) ± standard deviation in blue. First, it is observed that the averaged skewness remains, in any conditions, above 0.5 though it is admitted that a large population's distribution ($N \gg 0$) tends toward the Normal law, the pure gaussian, with a null value of skewness. This certainly results from the zero limit that is imposed by particles encounters. Now, when the particle density decreases from $10^4$ down to \textit{ca}. $10^3$ the skewness increases along with a much larger spreading of the values as it is shown from the sd increases (blue bars in Fig. 1). As a whole, this indicates that outsiders, the so-called microstates, would be more present, relatively to the particle numbers, when the density is low since the trend exponentially increases  at very low density, namely N=30 in this simulation: at very low density, skewness exceeds the value of 1 which is considered as a very asymmetric distribution.

\begin{figure}[H]
\centering
\includegraphics[width=0.8\textwidth]{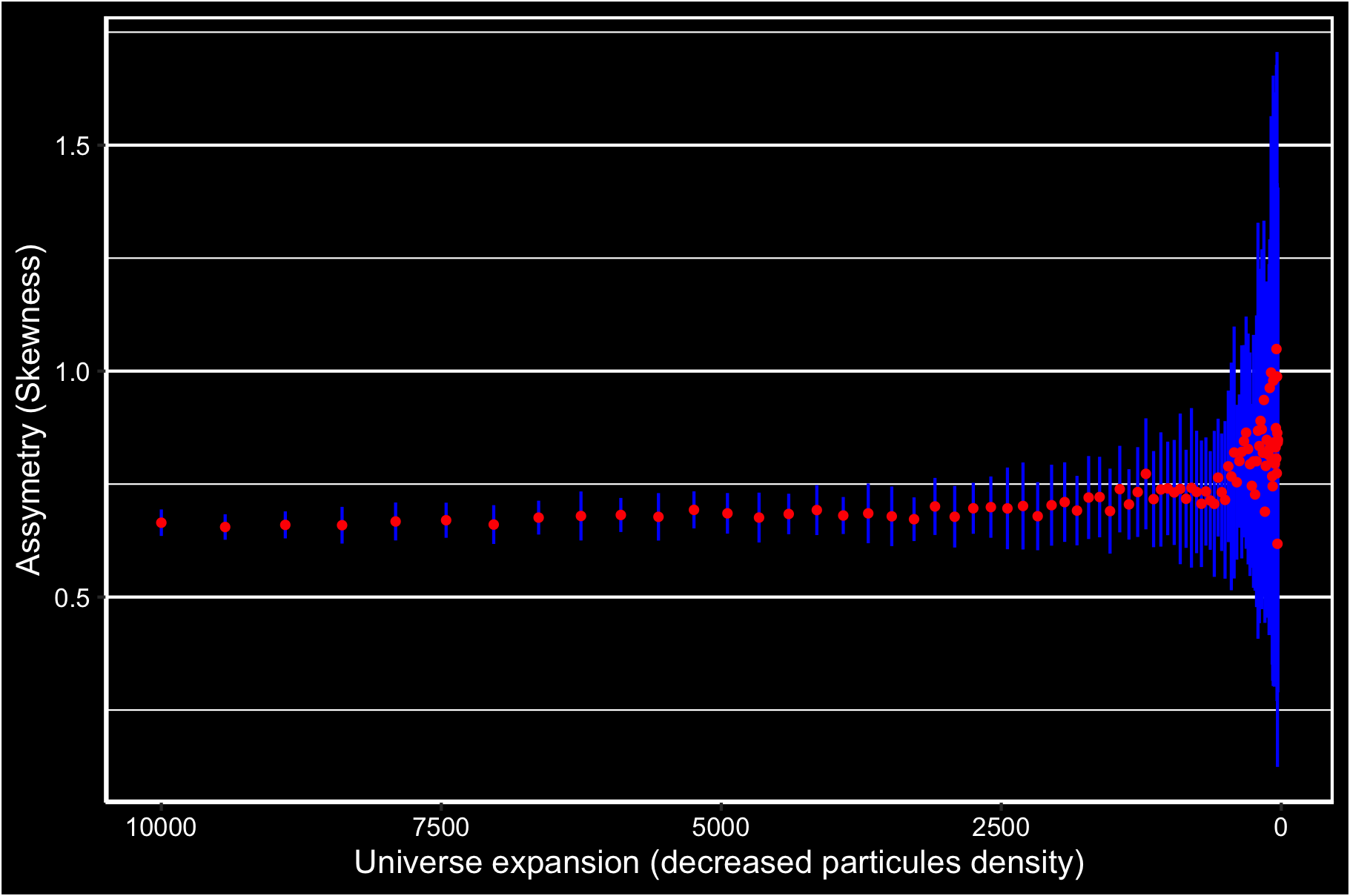}
\caption{Skewness evolution as the particles density decreases}
\end{figure}

To better exemplify the evolution of asymmetry as the particles density varies, Figure 2. display the spatial distributions and nearest-neighbors distances histograms for random simulations, with $32 \leq N \leq 4096$. Density curves (in red on Fig.2b) help to better visualize asymmetries of the distributions. As an example, the skewness increases from 0.67 up to 1.11 when the number of particles N decreases from 1 000 to 100. It is recalled that, here, the parameter considered is the nearest distances between particles. In Fig. 2a, the colors and size of the particles encode their neighbor's nearest distance: the smaller the size and the more red the color, the more close distance between another particle, the larger the size and the more blue, the more isolated ones.
\begin{figure}[H]
  \centering
  \begin{subfigure}[b]{0.45\textwidth}
    \includegraphics[width=\textwidth]{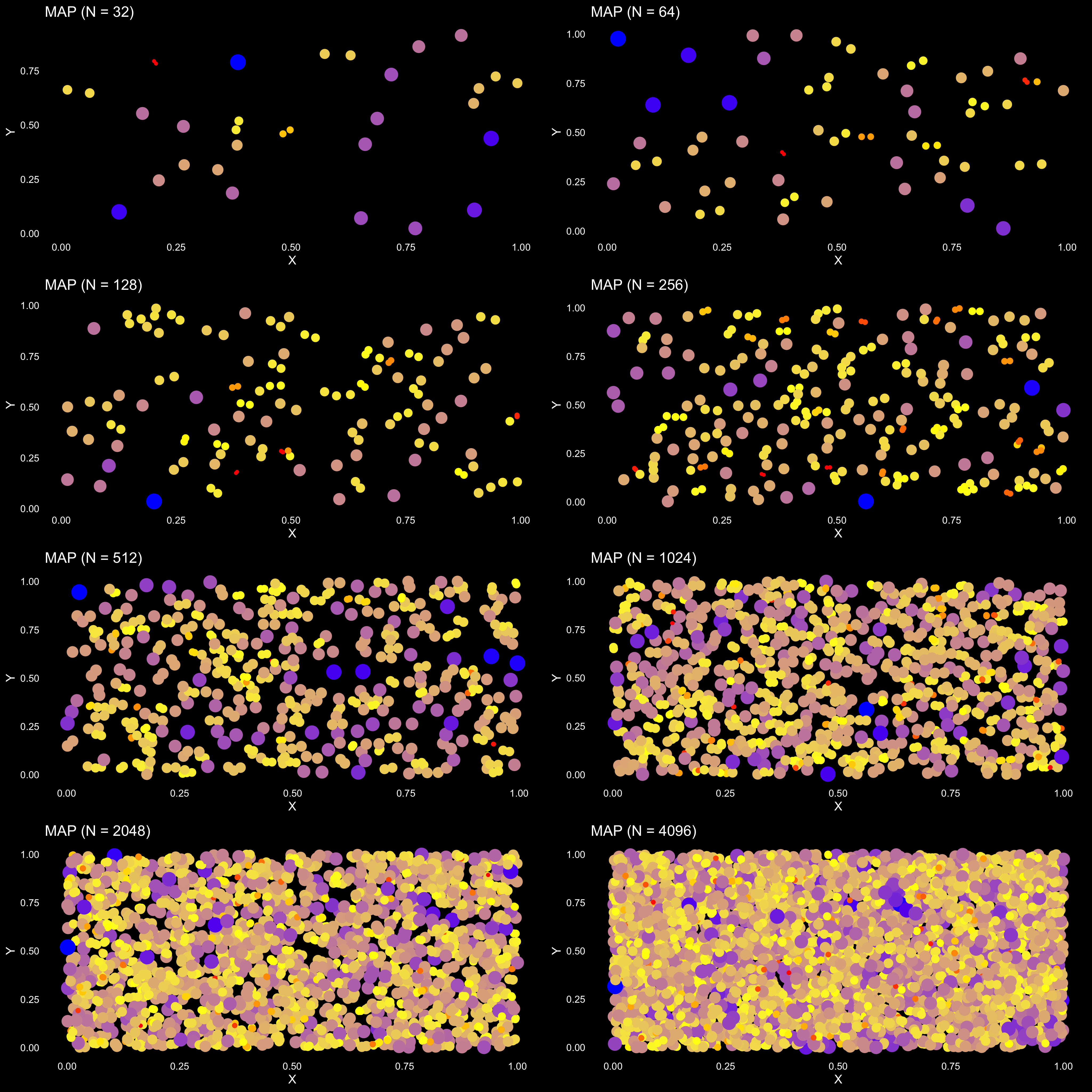}
    \caption{Random distributions}
    \label{fig:sub1}
  \end{subfigure}
  \hspace{0.5cm}
  \begin{subfigure}[b]{0.45\textwidth}
    \includegraphics[width=\textwidth]{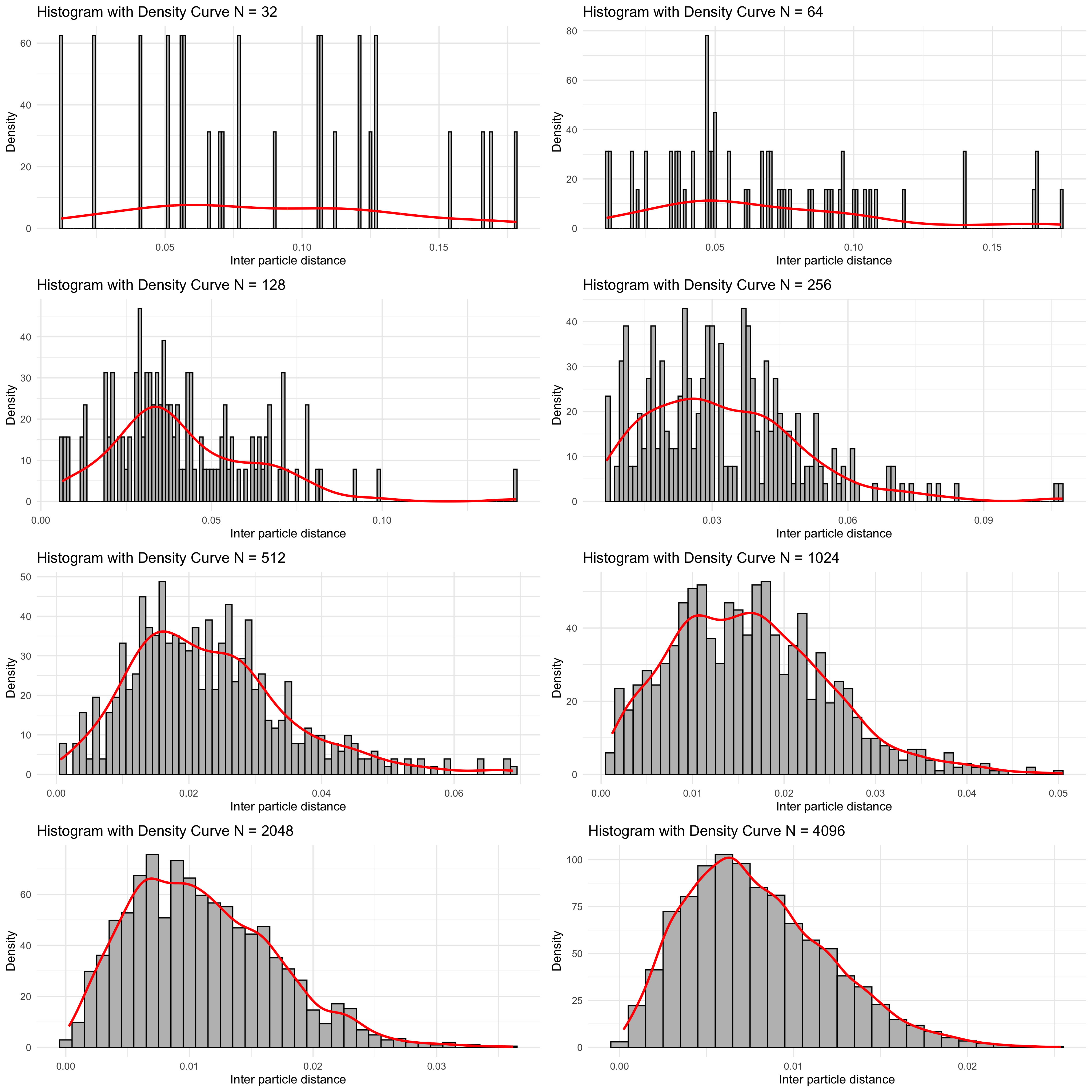}
    \caption{Skewness of distributions}
    \label{fig:sub2}
  \end{subfigure}
  \caption{Initial distributions ($32 \leq N \leq 4096$)}
  \label{fig:main}
\end{figure}

To better characterize the long tail of the nearest-neighbors distribution, a case with \( N = 4096 \) and \( n = 8192 \) was computed, yielding a skewness of 0.686. Attempts were made to assess whether this distribution exhibits a heavy tail and if this tail conforms to a Pareto approximation. For this purpose, the data were analyzed using the \texttt{extremefit} R package \cite{Exfit}. This package first identifies the threshold \( \tau \) beyond which the data may follow a Pareto distribution. Subsequently, data points above this extreme quantile are tested for adherence to a Pareto model. The \texttt{hill.adapt} function yields a value of 0.0139 (\texttt{HA\$Xadapt}) that is the adaptive threshold value (denoted as \( \tau \) in the literature) from which point the procedure estimates that the tail of the present data distribution can be reasonably modeled by a Pareto law.

\begin{figure}[H]
  \centering
  \begin{subfigure}[b]{0.8\textwidth}
    \includegraphics[width=\textwidth]{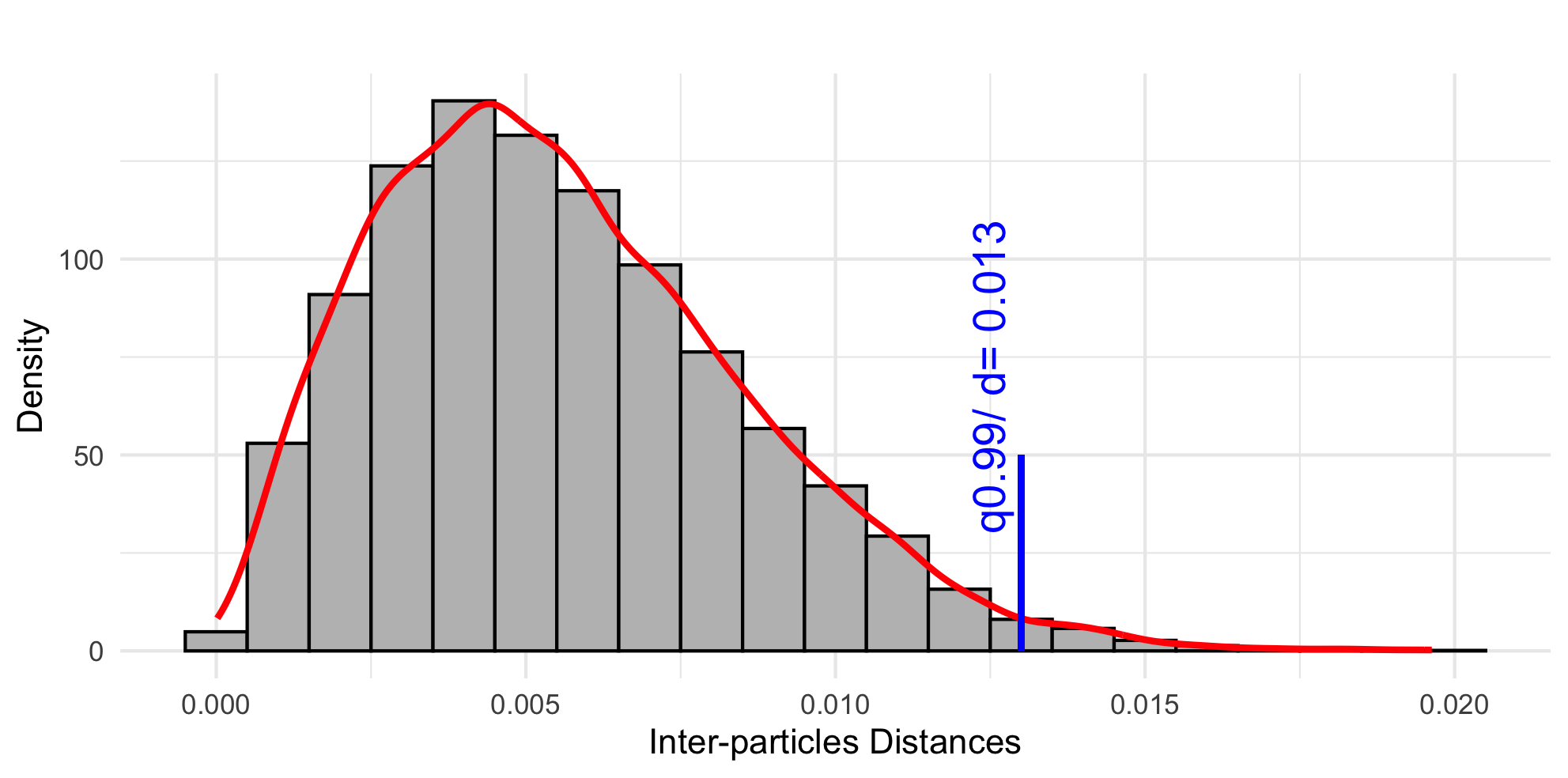}
    \caption{Nearest distances distribution}
    \label{fig:sub31}
  \end{subfigure}
  \hspace{0.15cm}
  \begin{subfigure}[b]{0.8\textwidth}
    \includegraphics[width=\textwidth]{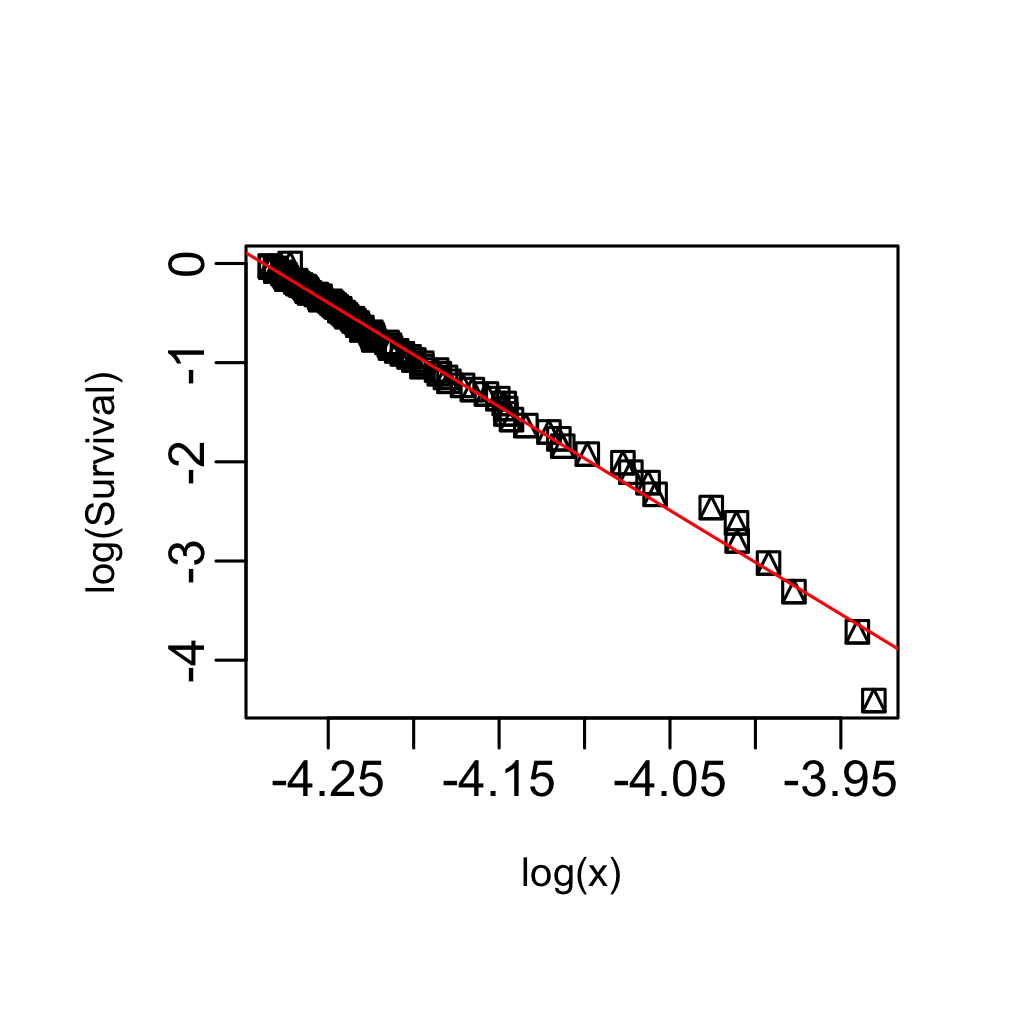}
    \caption{Log Log plot for the 0.99 extreme quantile}
    \label{fig:sub32}
  \end{subfigure}
  \caption{Pareto Evaluation for $N = 4096$ and $n = 8192$}
  \label{fig:main}
\end{figure}

Accordingly, the Fig. 3a displays the 0.99 quantile of this distribution, indicating that 1\% of data points (~41) exceed the 0.013 distance threshold. Above this distance, the distribution is better fitted by a power law (Fig. 3b) than by an exponential distribution. This observation aligns with the presence of a Pareto-distributed tail. In conclusion, such a distribution exhibits rare but statistically significant events that deserve attention. Extreme points in a Pareto tail are distinguished by their unusually high frequency compared to thin-tailed distributions, their scale-invariant (self-similar) structure, and their statistical modeling via heavy-tailed laws such as the Fréchet or generalized Pareto distributions. These properties imply a non-negligible probability of rare but catastrophic \footnote{in the René Thom meaning} events, making Pareto tails essential for accurate risk assessment in fields such as finance, environmental science, and engineering.

A second attempt was performed to allow quantizing entropy changes over time. Accordingly, a new model was designed  based on a 33x32 matrix  which n= 1024 cells may host 0 to N particles. As above, an initial random, so non-uniform,  distribution was set for N particles ($32 \leq N \leq 4096$). Then this initial distribution was allowed to evolve (at N constant) over few (8) time steps with a simple rule purposely mimicking gravity: the probability that a particular cell hires more particles at each step is weighted by the square of particles yet hosted at the previous one. It yields a more or less fast particle's clustering that depends on the initial partition and on the N/n particles/cells ratio.
Here, the thermodynamic entropy calculation equations follow : 
for a 32x32 matrix with N particles non-uniformly distributed , we use the Boltzmann entropy formula with Stirling's approximation. 
 
 1. Stirling's Approximation for a large number of particles $ n $:
\[
\ln(n!) \approx n \ln(n) - n
\]
2. Total Entropy of the System : the entropy $S$ is proportional to the logarithm of the number of microstates $\Omega$:
\[
S = k_B \ln(\Omega)
\]
where
\[
\Omega = \frac{N!}{\prod_{i=1}^{1024} n_i!}
\]
with:
\begin{itemize}
  \item $N = \sum_{i=1}^{1024} n_i$: total number of particles in the matrix,
  \item $n_i$: number of particles in cell $i$.
\end{itemize}

Using Stirling’s approximation, the entropy becomes:
\[
S = N \ln(N) - N - \sum_{i=1}^{1024} \left[ n_i \ln(n_i) - n_i \right]
\]
Simplifying (since $N = \sum n_i$):
\[
S = N \ln(N) - \sum_{i=1}^{1024} n_i \ln(n_i)
\]
 
So a 	high entropy (e.g., uniform distribution) indicates that particles are spread evenly (maximum disorder) and a low entropy (e.g., concentrated distribution) that particles cluster in few cells (high order). The equations quantify how much entropy decreases as particles cluster, reflecting how order increases in the corresponding system. Fig. 4 and 5 display  both how particle's spatial distributions and the resulting entropy evolve over time for N=512 and N=4096, respectively.

\begin{figure}[H]
  \centering
  \begin{subfigure}[b]{0.45\textwidth}
    \includegraphics[width=\textwidth]{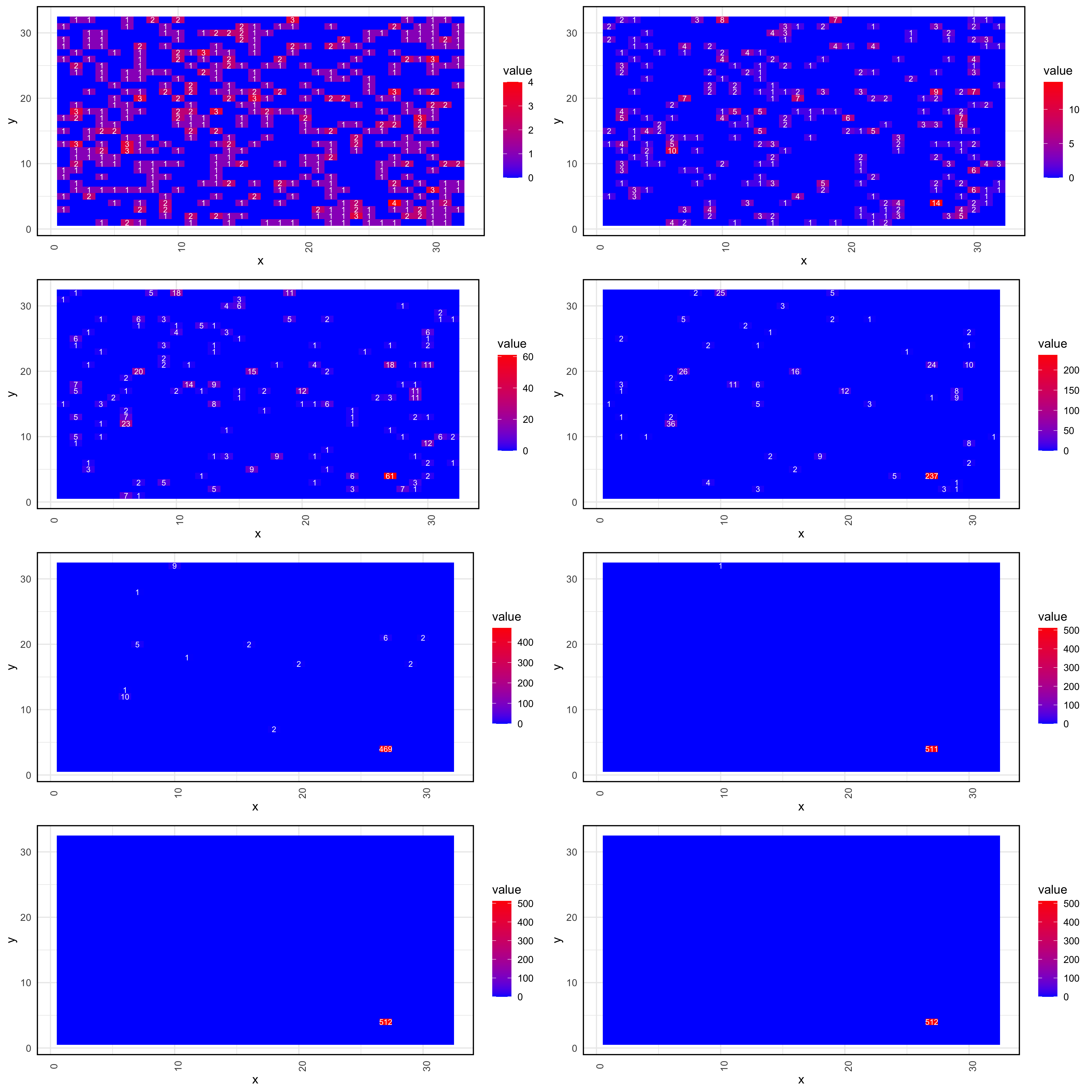}
    \caption{Particles distribution}
    \label{fig:sub1}
  \end{subfigure}
  \hspace{1cm}
  \begin{subfigure}[b]{0.45\textwidth}
    \includegraphics[width=\textwidth]{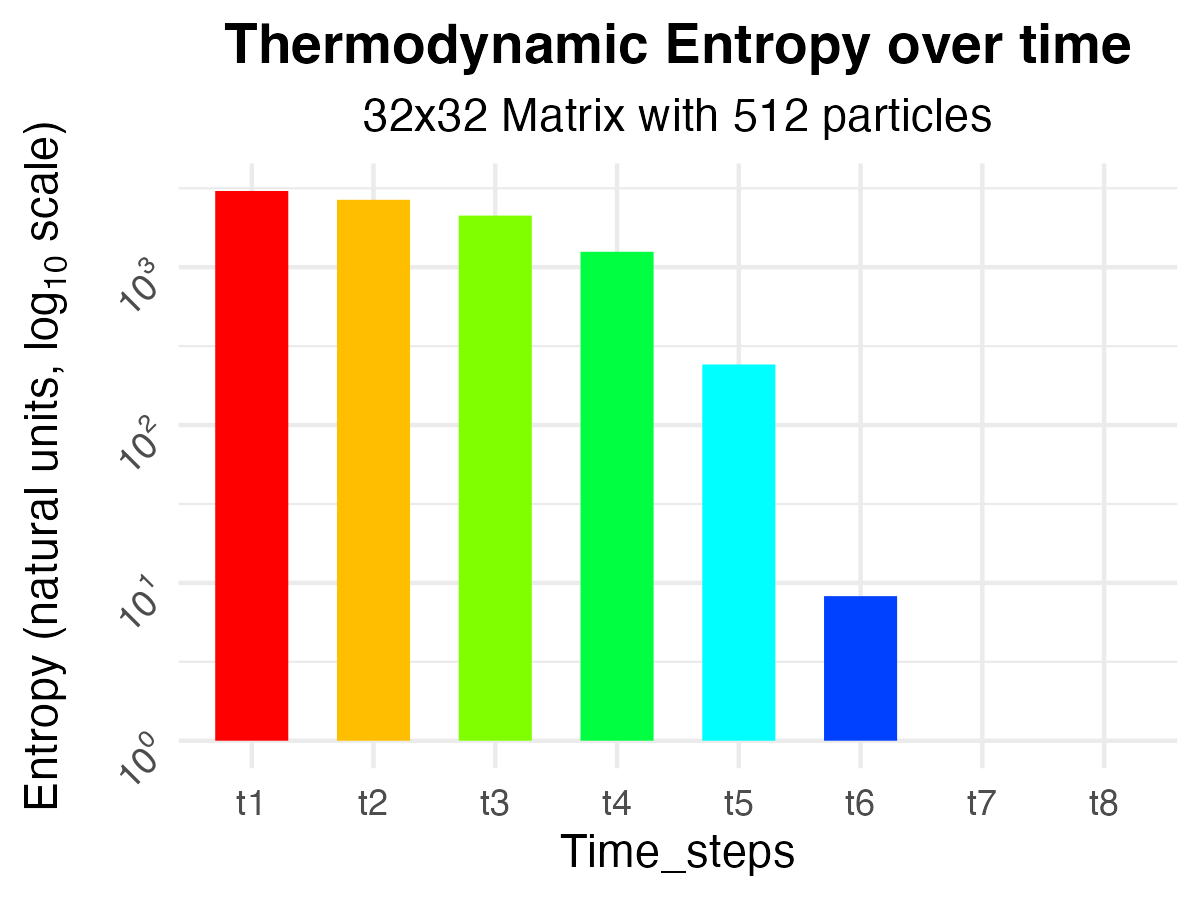}
    \caption{Entropy evolution over time }
    \label{fig:sub2}
  \end{subfigure}
  \caption{$\frac{N}{n} = 0.5$ Model}
  \label{fig:main}
\end{figure}

\begin{figure}[H]
  \centering
  \begin{subfigure}[b]{0.45\textwidth}
    \includegraphics[width=\textwidth]{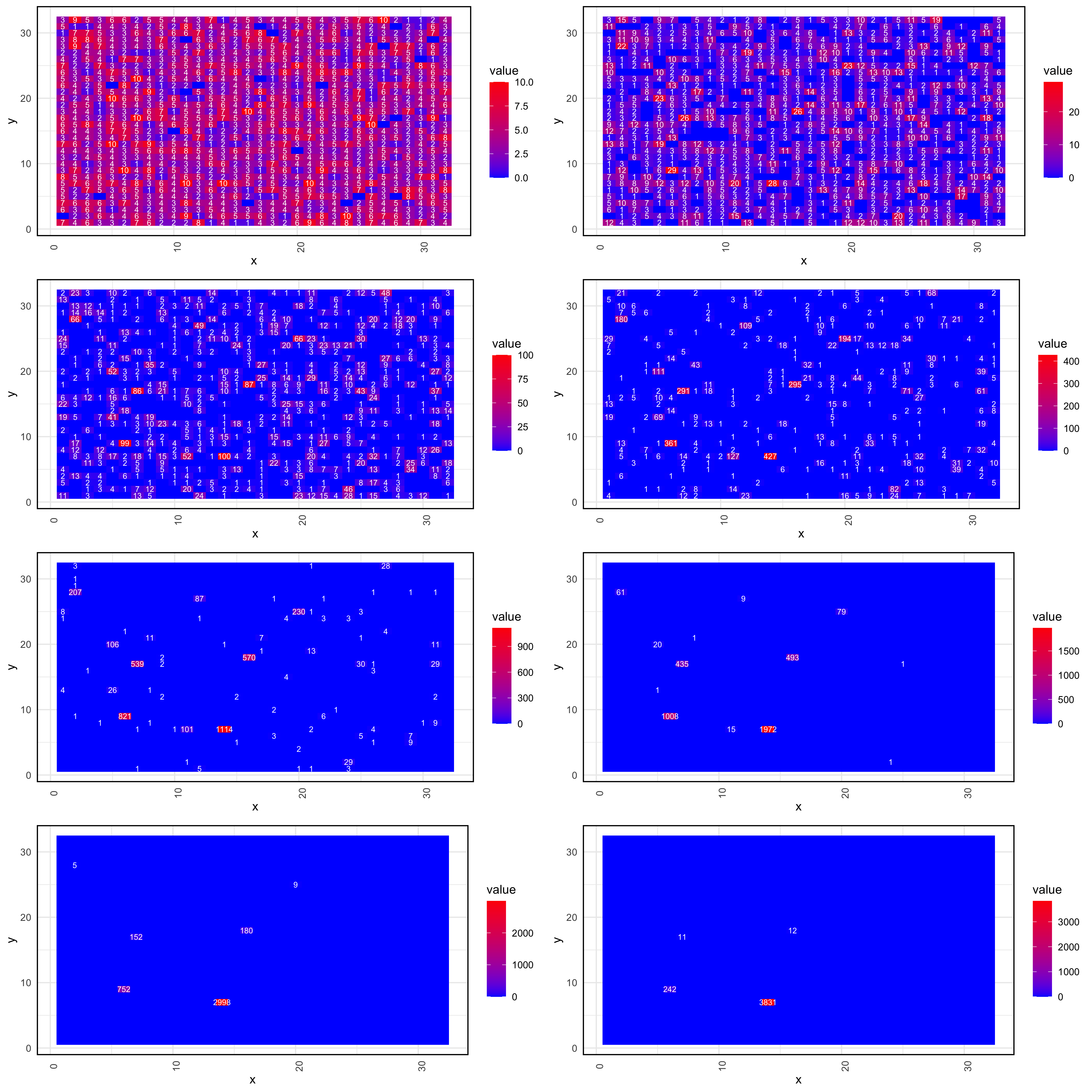}
    \caption{Particles distribution}
    \label{fig:sub1}
  \end{subfigure}
  \hspace{1cm}
  \begin{subfigure}[b]{0.45\textwidth}
    \includegraphics[width=\textwidth]{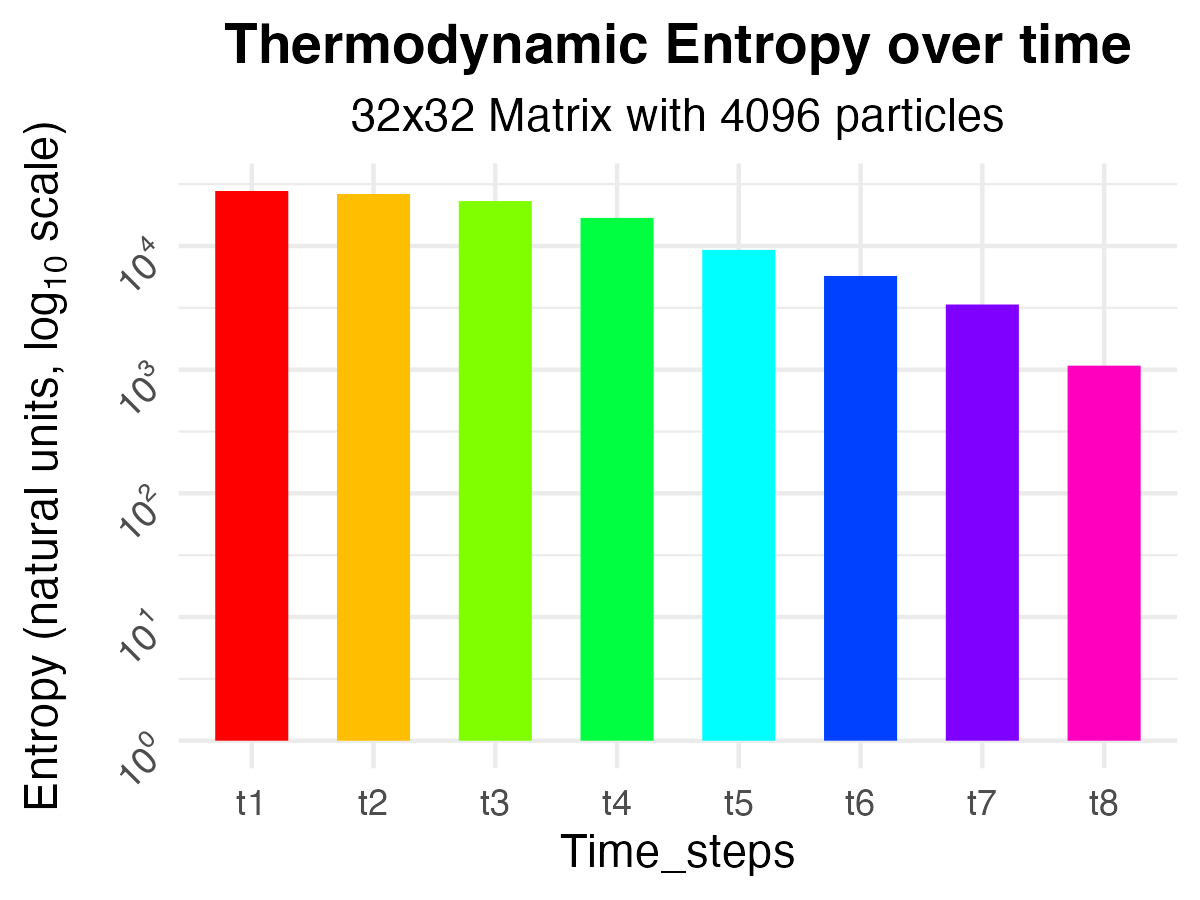}
    \caption{Entropy evolution over time }
    \label{fig:sub2}
  \end{subfigure}
  \caption{$\frac{N}{n} = 4$ Model}
  \label{fig:main}
\end{figure}

Figures 4a and 5a displays how many particles are hosted in each cell over time through 8 time steps. Figures 4b and 5b display the corresponding entropy evolutions over time on a log scale. A zero value, as shown at t7 ant t8  for N=512, indicates that every particle has migrated towards a single cell. It is observed that, in crowded conditions (N=4096), the clustering is slower and therefore the decrease in entropy proportionally weaker, as compared to rarefied spaces (N=512) where particles 'rapidly' cluster into a  single cell which, hence, originates a tiny, rare, low entropy world.
One must recall that, here, fast and slow are purely qualitative terms, as are the time steps that do not feature any particular delay.
If now we focus on the fate of an initially crowed (N=4096) piece of the vast universe, it arises that it evolve towards four little worlds which size span 3 log (Fig. 6). Actually, the smallest once hosts 11 particles while the largest hosts nearly 4 000 ones.
\begin{figure}[H]
\centering
\includegraphics[width=0.8\textwidth]{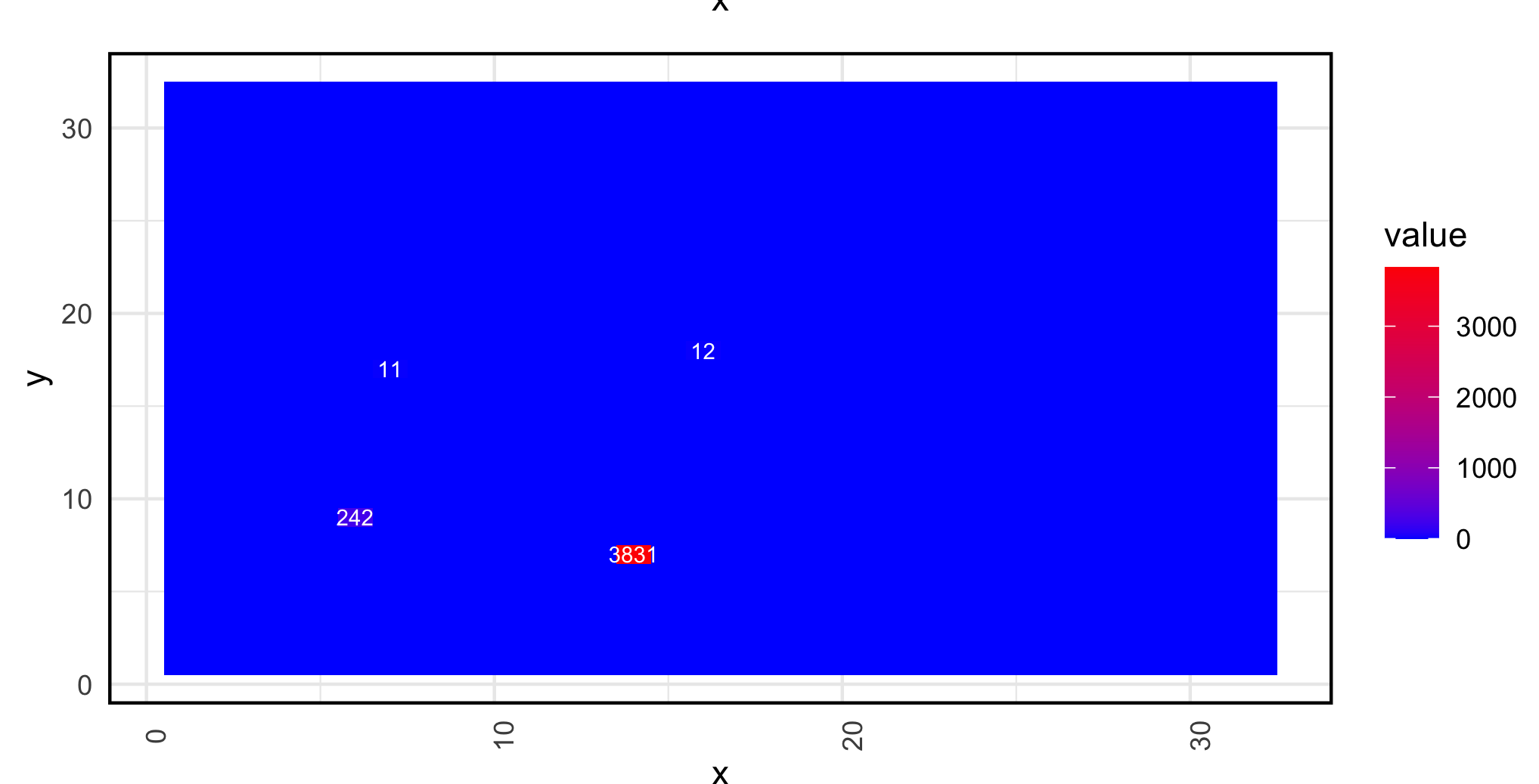}
\caption{Particles Distribution at t=8 for N=4096}
\end{figure}
 
It is likely that these four "worlds" will now evolve differently due to distances and inner particle interactions, which must depend on their respective size. It is tempting to see here a kind of fractal property that links the behavior of a single particle to that of a larger entity, i.e., one composed of a larger  number of individuals. The crucial point is that it is the interactions between elemental, constitutive entities that actually drive the dynamics and determine the future of such systems.

\section{Discussion}
Do outsiders worth ? There is the question this short note addresses. Usually in sciences, what attract one's eye in any observed distribution are its first and second moments, i.e. the mean and the standard deviation.  Tails, either left or right are often disregarded and simply assigned to some incertitudes on the measurements or as  individuals for which paying attention to does not matter. An exception could be security services which actively zoom on rare, unexpected, human behaviors spotted by control cameras : the "exception" is the algorithm target ! The point of view would likely change if now we question how these "outsiders" evolve over time as compared to those laying, let say, in the $\bar{x} \pm 2\sigma$ range. Do mainstreams actually rule the world ?
In this simulation, particles may feature a simple gas molecule or a small world, small regarding the system's dimension which is overwhelmingly larger than the "particle" one's. At any time scale, gravity forces or any other kind of interactions  warrant for observed particles spatial positions, whether they collide or not. As mentioned above, any distribution observed at a particular \textit{t} time serves as a starting point for a particular dynamics, its fate. It is known that very close initial conditions may result in further drastic divergences that can be characterized by the Lyapunov exponent.
Since this approach is derived from asymmetric distributions, even when $N \gg 0$ that is \textit{a priori} unexpected, we have to consider both \textit{"left"} and \textit{"right"} tails of the density curves. Since, here, skewness is always positive and > 0.5, the asymmetry is larger on the \textit{"right"} tail than on the \textit{"left"}  one. It results that isolated (large inter-particles distances) are more populated than clumps ones. That's what we shall describe as the \textit{"right"}  and \textit{"left"} cases.
So the \textit{"right"}  case features isolated particles or lone tiny worlds, the so-called  \textit{heavy tail}. This means that they will be not exposed to external forces, contributions or noise (think to solute/solvent interactions) and that they may evolve on their own, meaning on the sole basis of their internal forces\footnote{Here we consider that a particle may feature a small world on its own, submitted to its own laws, regardless its nature}. In opposite the \textit{"left"} case, though even less populated, do exist (see Fig. 2a), the more in a hot Universe. In opposite to the former, these particles/tiny worlds are submitted to external forces exerted by their close neighbors: they act upon, and are submitted to their nearest neighbors. This status is completely distinct from the isolated particles/worlds. So we have to admit that these two kinds of microstates must exhibit very distinct dynamics, over time to anchor the concept. 
These two status and induced-dynamics are akin, in my opinion, to what is called the \textit{conformational drift} in recombining proteins \cite{Drift}. These proteins are synthesized by a bacterium or a yeast after their genome has been purposely modified (GMO\footnote{Genetically Modified Organisms}). More precisely, dimeric proteins that are made from two distinct polymeric chains which assemble together once synthesized and thereby become biologically functional. If the recombinant protein chains are synthesized synchronously, they are able to properly assemble. On the opposite, if some delay exists between the synthesis of the two chains, the first one will start to relax: its conformation evolves due to interactions with the solvent molecules, its surrounding neighbors. As a consequence, the later synthesized second chain  will no more "recognize" its partner and the functional dimer will never be formed. This is what is referred as a \textit{conformational drift}. Such proteins misfoldings originate many pathologies as cystic fibrosis, among others.  Coming back to the microstates, it is very likely that they evolved distinctly depending on whether they do either lonely (the \textit{"right"} case) or in a crowded environment (the \textit{"left"} case) . 

As a tentative conclusion, these distinct microstates may coexist and evolve within the macrostate without violating the probabilistic interpretation of the second law of thermodynamics. While they evolve toward different entropy values, their contributions to the global macrostate are negligible due to their scarcity and minuscule scale. The time evolution of any system is governed by its interaction network: even a minor initial discrepancy can lead to systemic divergences over time, appearing "alien" relative to the bulk. However, observed divergences ultimately stem from the inherent stability of dynamical systems. If such microscopic worlds were to seek analogous entities, they would require immense patience and the development of extraordinarily powerful, long-range radio telescopes.
\hspace{1cm}

The author greatly acknowledges Dr. Salim LARDJANE  advices for the conception of the entropy driven model and Pr. Gilles DURRIEU for the Pareto analysis, both from LMBA (UMR CNRS 6205).

% \appendix
% \includepdf[pages=-]{microRcode.pdf}

\end{document}